\documentclass[apj]{emulateapj}
\usepackage[varg]{txfonts}
\usepackage{graphicx}
\usepackage{float}
\usepackage{natbib}
\usepackage[english]{babel}
\usepackage{color}
\usepackage{soul}

\submitted{}
\received{}
\accepted{}

\slugcomment{Submitted to AJ}
\shorttitle{An M dwarf Companion to an F-type Star}
\shortauthors{Eigm\"uller et al.}

\begin{document} 
\newcommand{\Mass}[0]{1.493}
\newcommand{\Masse}[0]{0.073}
\newcommand{\Radius}[0]{1.474}
\newcommand{\Radiuse}[0]{0.040}
\newcommand{\Masst}[0]{0.188}
\newcommand{\Masste}[0]{0.014}
\newcommand{\Radiust}[0]{0.234}
\newcommand{\Radiuste}[0]{0.009}

\title{An M dwarf Companion to an F-type Star in a young
main-sequence binary}


\author{Ph. Eigm\"uller\altaffilmark{1,}\altaffilmark{2}, 
J. Eisl\"offel\altaffilmark{2},
Sz. Csizmadia\altaffilmark{1},
H. Lehmann\altaffilmark{2},
A. Erikson\altaffilmark{1},
M. Fridlund\altaffilmark{1,}\altaffilmark{3,}\altaffilmark{4,},
M. Hartmann\altaffilmark{2},
A. Hatzes\altaffilmark{2},
Th. Pasternacki\altaffilmark{1},
H. Rauer\altaffilmark{1,}\altaffilmark{5},
A. Tkachenko\altaffilmark{6,}\altaffilmark{7},
H. Voss\altaffilmark{8}}

\altaffiltext{1}{Institute of Planetary Research, German Aerospace Center,
Rutherfordstr. 2, 12489 Berlin, Germany}
\altaffiltext{2}{Th\"uringer Landessternwarte Tautenburg, Sternwarte 5, 07778
Tautenburg, Germany}
\altaffiltext{3}{Department of Earth and Space Sciences, Chalmers University of
Technology, Onsala Space Observatory, 439 92, Onsala, Sweden}
\altaffiltext{4}{Leiden Observatory, University of Leiden,PO Box 9513, 2300 RA, Leiden, The Netherlands} 
\altaffiltext{5}{Department of Astronomy and Astrophysics, Berlin University of
Technology, Hardenbergstr. 36, 10623, Berlin, Germany}
\altaffiltext{6}{Instituut voor Sterrenkunde, KU Leuven, Celestijnenlaan 200D, 3001 Leuven, Belgium}
\altaffiltext{7}{Postdoctoral Fellow of the Fund for Scientific Research (FWO),
Flanders, Belgium}
\altaffiltext{8}{Universitat de Barcelona, Department of Astronomy and Meteorology, Mart\'i i Franqu\`es, 1, 08028 Barcelona, Spain}


\begin{abstract}
Only a few well characterized very low-mass M dwarfs are known today. 
Our understanding of M dwarfs is vital 
as these are the most common stars in our solar neighborhood.
We aim to characterize the properties of a rare F+dM stellar system for a better
understanding of the low-mass end of the Hertzsprung-Russel diagram. 
We used photometric light curves and radial velocity follow-up measurements to
study the binary. 
Spectroscopic analysis was used in combination with isochrone fitting to
characterize the primary star. 
The primary star is an early F-type main-sequence star with a mass of 
(\Mass{} $\pm$ \Masse{}) $M_\odot$ and a radius of 
(\Radius{} $\pm$ \Radiuse{}) $R_\odot$. 
The companion is an M dwarf with a mass of  
(\Masst{} $\pm$ \Masste{}) $M_\odot$ and a
radius of 
(\Radiust{} $\pm$ \Radiuste{}) $R_\odot$. 
The orbital period is 
$(1.35121 \pm 0.00001)d$. The secondary star is among the lowest-mass M
dwarfs known to date. 
The binary has not reached a 1:1 spin-orbit synchronization.
This indicates a  young main-sequence binary with an age below
$\sim$250\,Myrs. 
The mass-radius relation of  both components are in agreement with this finding.
\end{abstract}



\section{Introduction}
\label{sect:intro}
Understanding stellar evolution requires a knowledge, to high precision, 
of the fundamental parameters of stars in different stages of their evolution.
The study of detached eclipsing binaries offers us a unique method of
determining the bulk parameters of stars and to compare
these measurements to the predictions from stellar models.
Stellar models succeed in predicting the mass-radius relation to an accuracy of a few percent for main-sequence 
stars with $M_{\sun}<M_{\star}<5\cdot M_{\sun}$ \citep[e.g.][]{andersen1991}. 
Systematic discrepancies between model and observation in the mass-radius relation for a given age
 have been associated with the amount of convective core overshoot by  \citet{clausen2010}, but these are below 1\%.
Low-mass stars with $M_{\star}<M_{\sun}$ are the most common stars in the solar neighborhood, but only a
very limited number of these are well-characterized  \citep{torres2013}.
For these stars, stellar models also show systematic discrepancies in
the observed mass-radius
relations, but on a larger scale.
Over 30 eclipsing very low-mass stars (VLMSs) with masses below $0.3M_\sun$
and radii known to better than 10\% have been observed so far \citep[e.g.][]{parsons2012, pyrzas2012, nefs2013, gomez2014, zhou2014, kraus2015, david2016}.
However, only eight have radii known to  a precision better than
2\%.
Additionally, a few VLMSs have been characterized by interferometric
observations  \citep{lane2001, segransan2003, berger2006, belle2009, demory2009,
boyajian2012} with accuracies up to a few percent.

When evaluating detached eclipsing binaries (DEBs) and single star observations,
the highest discrepancies between models and observations have been found for
stars with masses
between $0.3 M_\sun < M_\star < 1M_\sun$ which are not fully
convective  \citep[e.g.][]{lopez2007, ribas2006, boyajian2015}.
For VLMSs with masses below $0.3M_\sun$, which have a fully
convective interior,
current models seem to systematically underestimate the radii by up to 5\%
percent compared to observations of detached
binaries  \citep[e.g.][]{torres2010, boyajian2012, spada2013, mann2015}. 
Interferometric radius determinations of single VLMSs show even larger
discrepancies to the models for some stars  \citep{boyajian2012,
spada2013}, but in general agree with the above findings.
Currently there is no satisfying explanation for the discrepancy between models and observed radius estimates.
\citet{mann2015} characterized a large set of low-mass stars using
spectrometric observations. They found similar discrepancies
to the stellar similar to what was seen
in the sample of characterized DEBs.
Using data from over 180 stars they confirmed that stellar models tend 
to underestimate stellar radii by
$\sim5\%$ and overestimate effective temperatures by $\sim 2.2\%$.
Although a large influence of metallicity on the $R_\star-T_{eff}$ correlation was found,
neither this correlation nor any other could explain the observed discrepancies to current stellar models.

All
state-of-the-art stellar evolution models  \citep[e.g.][]{baraffe1998,
dotter2008, bressan2012} give comparable mass-radius relations for stars
with masses below $0.7 M_\sun$ and older than a few hundred Myrs.  
The differences among various stellar evolution models are
well below a few percent. 

On the other hand, for young main-sequence
VLMSs with ages well below 250\,Myrs,  the differences between the models
are much larger.  
Older low-mass stars require a  precision better than 2\% in the bulk parameters
in order to  test stellar evolution  models  \citep{torres2013},  
but with young systems it is sufficient to characterize these
with a much lower precision.
This makes young main-sequence objects ideal for testing stellar evolution
models. 
Unfortunately the number of known young main-sequence low-mass stars is
very limited. Recently two such young systems with ages below $\sim$10\,Myrs have been characterized \citep{kraus2015,david2016}. 

Ages of main-sequence stars are estimated by different methods. Besides using stellar evolution models which
correlate basic observables (e.g. mass, radius, luminosity, and temperature)
with the age of the star, gyrochronology allows one
to correlate the rotational period and color index with the stellar age of
cool stars  \citep[e.g.][]{barnes2010}.
For close binaries this method is limited by dynamical
interactions that might have influenced the rotational period of the stars.
For stars with uninterrupted high precision photometric observations we can use  
asteroseismology to determine the age of a star
 \citep[e.g.][]{aerts2010}.
The accuracy of the age determination with gyrochronology is $\sim 10\%$  \citep{delorme2011}.
The ages determined with different stellar model can deviate by
$\sim10\%$ for young stars and from 50\% up to 100\% for older stars  \citep{lebreton2014a}. 
Only asteroseismology in combination with stellar evolution models can provide the age of main-sequence stars with an accuracy better than 10\%  \citep{lebreton2014b}.
If the observed system is a cluster member, the age of the star can also be inferred from the age of the cluster.
For close binary stars whose orbits  are not yet synchronized, the upper limit of the age of
the system might also be given by the time scale of synchronization \citep[e.g.][]{drake1998}.

We present a possibly young F+dM SB1 binary system with a short orbital
period and a low eccentricity. 
We characterize the system and both components using photometric and spectroscopic data.
To characterize the primary star we use spectral analysis 
and compare the results to stellar evolution models. 
We model the light curve of the primary eclipse and in combination with the radial velocity measurements determine the mass-radius relation of the low mass companion.
This enables us us to
estimate  an upper limit for the age of the unsynchronized system.  

\section{Observations}
\subsection{Photometric Observations}
\label{sect:phot}
Photometric observations were taken during surveys for transiting planets with
the \textit{Berlin Exoplanet Search Telescope}  \citep[BEST;][]{rauer2004} 
and the \textit{Tautenburg Exoplanet Search Telescope}
 \citep[TEST;][]{eigmueller2009}. With both telescopes the same circumpolar
field 
close to the galactic plane was observed for several years. 
Technical details on the surveys are given in Table \ref{table::tele}.
For both surveys typically between a few tens of thousands up to a hundred thousand stars have been observed
simultaneously within the field of view.
In Table \ref{table::photobs} the observing hours per year for this field
are listed.

\begin{table}
\caption{Technical parameters of the BEST and TEST surveys.}
\centering
\label{table::tele}
\begin{tabular}{lcccc}
\hline \hline
 & BEST Survey &  TEST Survey \\
\hline
Site & TLS (2001 - 2003) & TLS\\
 & OHP (2005-2006) & \\
Aperture & 200~mm & 300~mm\\
Camera & AP 10 & AP16E\\
Focal ratio & f/2.7 & f/3.2\\
Pixel scale & 5.5 arcsec/pixel & 1.9 arcsec/pixel\\
Field of view & 3.1$^{\circ}$x3.1$^{\circ}$ & 2.2$^{\circ}$x2.2$^{\circ}$\\
Readout Time & $\sim90s$ & $\sim30s$\\
Exposure Time & 240s & 120s\\
No. of frames on target & 800 & 6000\\
\hline
\end{tabular}
\end{table}

The eclipsing binary presented in our work was detected in both surveys
 \citep{voss2006, eigmueller2012} as planetary candidate.
The object was published as an uncharacterized Algol type binary in
 \citet{pasternacki2011} 
with the identifier \objectname[UCAC4 714-021661]{BEST F2\_06375} after its
planetary status was excluded.
First estimates of the mass-radius relation gave hints on a possibly
inflated very low mass star, which led to further follow-up observations.
 
The observations with the BEST were taken between 2001 and 2006, with a
relocation of the BEST in 2003/2004 from the 
Th\"uringer Landessternwarte Tautenburg (TLS) in mid-Germany to the Observatoire de 
Haute Provence (OHP) in southern France.  
The survey with the TEST was carried out between 2008 and 2011 at TLS. 
Over 250 hours of photometric data were gathered between 2001 and 2011 in nearly
100 nights with these two surveys (cf. Table \ref{table::photobs}). 
The standard deviation of the unbinned light curve is typically better than 10\,mmag. 

The data gathered with both telescopes were reduced and analyzed with the
pipelines designed for the respective instruments. 
The pipeline used for the TEST data is described in  \citet{eigmueller2009}. The
methods used to analyze the BEST data set 
have been applied to various published BEST data sets  \citep[e.g.][]{fruth2012,
fruth2013, klagyivik2013}. 
The data reduction included standard bias and dark subtraction as well as a
flat field correction. 
The detrending for both data sets was done using the sysrem algorithm
 \citep{tamuz2005}. Effects present in only a few thousands of stars 
have been corrected.
A detrending of the individual light curves was not performed.

\begin{figure}
\centering
\epsscale{1}
\plotone{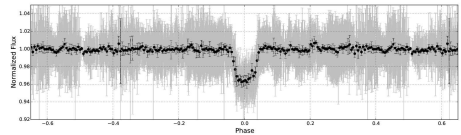}
\caption{The phase-folded light curve. Black points denote data binned to 10 minutes in phase,
while the gray points show the original data. Vertical lines show the uncertainties for single measurements.}
\label{fig::phaseall}
\end{figure}

For our study
we combined both data sets 
giving us a light curve with over 6800  data points (TEST: $\sim$6000,
BEST:$\sim$800). 
For the phase folded light curve we measure 
a standard deviation below 2\,mmag in the out-of-transit region
using values binned by up 10 minutes. 
The whole phase folded light curve is shown in Figure \ref{fig::phaseall}.

\begin{table}
\caption{List of the photometric observations of the eclipsing binary. For each
telescope the year, the number of observing nights, and the observing hours per year
are given.}
\centering
\label{table::photobs}
\begin{tabular}{lcc||lcc}
\hline \hline
& BEST & & & TEST & \\
Year & Nights  &  Observing   & Year & Nights  &  Observing   \\
   &     [\#]      &     hours &   &     [\#]      &     hours\\
\hline
 2001 & 3 & 3.8 & 2008 & 7 & 18.7\\
 2002 & 10& 18.6 & 2009 & 31& 95.9\\
 2005 & 4 & 10.0 & 2010 & 3 & 6.3\\
 2006 & 6 & 11.0 & 2011 & 34& 88.1\\
\hline
\end{tabular}
\end{table}

\subsection{Spectroscopic Observations}
\label{sect:spec}
Spectroscopic follow-up observations were performed with the
Tautenburg 2-m telescope using the Coud\'e-Echelle spectrograph
with an entrance slit that projected to 2'' on the sky. The observed wavelength
range covered
4700\AA\mbox{ } and 7400\AA\mbox{ } with a resolving power
($\lambda/\Delta\lambda$) of 32,000. For the wavelength calibration, 
spectra of a Thorium-Argon lamp were taken directly before and after the
observations. 
Stellar spectra were taken with exposure times of 1800 seconds which resulted in a typical S/N of 20-35.
In 2010  a few spectra of the binary system were taken between January and September to get an initial characterization of the transiting system.
In November/December 2012 additional spectra were obtained primarily  for radial velocity (RV) measurements needed to constrain the orbital motion. 
For the data reduction, standard tools
from IRAF were used including bias subtraction, flat-field correction, and
wavelength calibration. 
The RV  was determined using the IRAF \textit{rv} module.  
 
\section{System Parameters}
\label{sect:system}
The catalog information of the system is given in Table \ref{table::cat}.

\begin{table}
\caption{Catalog information of the eclipsing binary investigated here. Vmag
as given in UCAC4 catalog  \citep{zacharias2013}.}
\centering
\label{table::cat}
\begin{tabular}{l|c}
\hline \hline
 Parameter & Value   \\
\hline
Position & $02^{h} 40^{m} 51.5^s +52^d 45^m 07^s$\\
UCAC4 ID\footnote{ \citet{zacharias2013}} & UCAC4 714-021661\\
2MASS ID\footnote{ \citet{skrutskie2006}} & 02405152+5245066 \\
Bmag (UCAC4)& $12.287 \pm 0.02$\\
Vmag (UCAC4)& $11.769 \pm 0.02$\\
Jmag (2MASS)& $10.771 \pm 0.028$\\
Hmag (2MASS)& $10.618 \pm 0.032$\\
Kmag (2MASS)& $10.564 \pm 0.026$\\
pmRA (UCAC4)& $-1.7 \pm 0.8$ mas/yr\\
pmDE (UCAC4)& $-5.6 \pm 1.0$ mas/yr\\
\hline
\end{tabular}
\end{table}

\subsection{Modeling of the Photometric and Radial Velocity data}
\label{sect:lc}

\begin{figure}
\centering
\epsscale{1}
\plotone{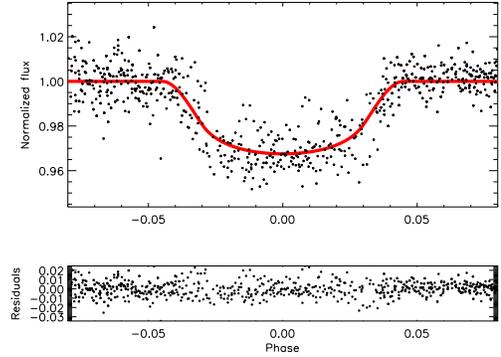}
\caption{The phase-folded light curve of the eclipse. Black points denote single
measurements, while the red line shows the best fit. }
\label{fig::phase}
\end{figure}

A simultaneous fit of the radial velocity and photometric data was performed.
The out-of-eclipse part of the light curve did not show any sign of ellipsoidal
variation at the level of precision of our observations 
(Figure \ref{fig::phaseall}).
Therefore we decided to use the spherical model of  \citet{mandel2002} for the
light
curve modeling.
The expected signal of the secondary transit would have an amplitude of $\sim$0.1\,mmag which would be undetected given our
red noise error of 2\,mmag.
To optimize the fit, we used a genetic algorithm  \citep{geem2001} to search 
for the best match between the observed and the modeled light curve. One
thousand individuals
were used in the  population and 300 generations were produced. 
The best fit found
by this procedure was further refined using  a simulated annealing chain
 \citep{kallrath2009}. 
The error was estimated using 
$10^4$ random models with  values within
$\chi^2+1\sigma$ of our best solution.
Figure \ref{fig::phase} shows
1 $\sigma$ error bars  \citep[for details of the code and implementation of
the algorithms see][]{csizmadia2011}.
For the light curve modeling we used the unbinned data. 
The effect of the exposure time was
taken into account by using a 4-point Simpson-integration 
\citep[e.g.][]{kipping2010}.

\begin{figure}
\centering
\epsscale{1}
\plotone{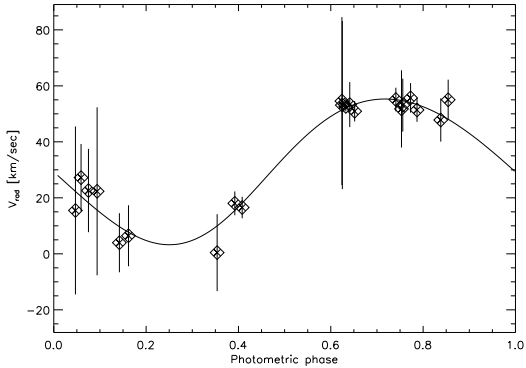}
\caption{Radial velocity measurements of the system.
The best fit gives a $\gamma$-velocity 
of (30.5 $\pm$ 0.5)\,km/s and a semi-amplitude of K = (26.1 $\pm$ 0.8)\,km/s.}
\label{fig::rv}
\end{figure}

Free parameters were the scaled semi-major axis ratio $a/R_s$, the inclination $i$, 
the radius ratio of the two stars $R_2 / R_1$, the epoch, 
the period, the $\gamma$-velocity, the semi amplitude of the radial velocity $K$,
the eccentricity $e$, the argument of periastron $\omega$, 
and the combination $u_+ = u_a + u_b$,  where $u_a$ and $u_b$ are the linear and
the quadratic limb
darkening coefficients of the quadratic limb darkening law. The parameter $u_-=u_a-u_b$
was fixed at the value found by interpolation of the R-band values of
 \citet{claret2011}. 
When we performed a fit using free limb darkening combinations as a check,
we got $u_-=+0.08\pm0.17$, compatible with the previous theoretical value. The 
other parameters were also within the error bars.
The results of the fit are presented in Table \ref{table::lcmodeling}.
Figure \ref{fig::phase} shows the phase-folded light curve over-plotted 
by the fit along with the residuals. Although 
the noise in single photometric measurements is 
large, the combined data allow us to reach a precision of $\sim$2\,mmag in 10
minute bins in the phase folded light curve.
The radial velocity data with the best fit are shown in Figure \ref{fig::rv}.\\

\begin{table}
\caption{Modeling Parameters. The given errors correspond to the
$1\sigma$ uncertainties. 
$a/R_1$; the impact parameter $b$ 
($b$ was calculated via 
$b= \frac{a\cdot(1-e^2)}{1+e \cdot cosv_0}*\sqrt{1-\sin^2i \cdot sin^2v_0}$
 where $v_0=90^\circ -\omega+\theta$ the mean anomaly at the mid-transit moment, see \citet{gimenez1983}); the inclination $i$ of the system; the radius
ratio $R_2/R_1$; the limb darkening coefficients $u^+$ and $u^-$; the
eccentricity $e$ of the system; the period $P$; the epoch of the system and the radial velocity semi amplitude $K$.}
\centering
\label{table::lcmodeling}
\begin{tabular}{l|c}
\hline \hline
Parameter & Value\\
\hline
$a/R_1$ &  $4.12 \pm 0.06$\\
b &  $0.45 \pm 0.03$\\
i &  $84.1^\circ \pm 0.3^\circ$\\
$R_2$/$R_1$ & $0.1601 \pm 0.0017$ \\
$u^+$ & $1.05 \pm 0.07$\\
$u^-$ & $-0.02$ \mbox{ } (fixed)\\
e & 0.070 $\pm$ 0.063\\
$\omega$ & $227^\circ \pm 13^\circ$\\
P & 1.35121d $\pm$  1$\cdot10^{-5}$d\\
Epoch &  2452196.1196 $\pm$ 0.0032 HJD\\
$\gamma-velocity$  & (30.50 $\pm$ 0.50) km/s\\
$K$  & (26.10 $\pm$ 0.76) km/s\\
\hline
\end{tabular}
\end{table}

\subsection{Stellar Parameters}
\label{sect:specclass}
To determine the atmospheric parameters of the primary component we created
a high quality spectrum by adding all the single observations
after applying an RV shift to account for the orbital motion. This resulted
in a co-added spectrum with S/N over 90. The analysis was performed
over the wavelength range 4740\AA\mbox{ } to 6400\AA.
using the GSSP program  \citep[Grid
search in Stellar Parameters;][]{lehmann2011, tkachenko2012}.

The normalization of the observed spectra during the reduction is 
difficult and the results strongly depend 
on the accuracy of the derived local continuum.
 We used the comparison of the co-added spectrum with the synthetic ones for an additional continuum correction.
 The analysis was done in three ways: a) without any correction, 
 b) by multiplying the observed spectrum by a factor calculated from a least squares fit between
 observed and synthetic spectrum, and c) with a re-normalization applied on smaller
 scales to get a better fit to the wings of the Balmer lines (mainly $H\beta$)
 and with regions excluded for which the analysis showed distinct deviations of
 the continuum from the calculated continua.
 Most of the atmospheric parameters obtained with the three different approaches agreed to within $1\sigma$. 
 However, approach c) gave a significantly higher value of the effective
temperature, $T_{eff}$ = 7350 $\pm$ 80 K, which differed by 
almost 2$\sigma$ from the
results of the other two methods. This demonstrates
the sensitivity of $T_{eff}$ caused by 
 small changes in the $H\beta$ wings.

The parameters
$T_{eff}$, $log\,g$, $v_{turb}$, $[Fe/H]$, and $v\, sin(i)$ and their errors
were derived using a grid.
Thus, the errors include all interdependencies between the  parameters.  All
other metal abundances and their errors were determined separately, fixing all
atmospheric parameters to their best fitting values.
The formal  $1\sigma$ error on $T_{eff}$ (80 K) based on error statistics
is probably too small due to systematic errors stemming from the continuum
normalization. We use a larger error that includes the systematic error 
introduced by this normalization.
 
As determining the stellar parameters is crucial and a possible source of
systematic errors in the characterization of the companion, the results have been
verified using another method described in  \citet{fridlund2010}. 
Stellar parameters of both methods are in agreement with each other. 
Only for $T_{eff}$ we found a larger uncertainty of $\pm 200K$. 
This error agrees with our previous finding that the normalization
of the spectrum can result in an underestimate in the error of the
effective temperature and thus the spectral classification.
In Table \ref{table::stellarmod} the results for the different approaches are given. 
For the estimates of mass and radius of the primary star we used the results from the GSSP 
approach with the small-scale re-normalization (c). For 
the error estimate of $T_{eff}$  
we used  250\,K  which corresponds to $\sim 3\sigma$ 
uncertainty in approach (c).

\begin{table}
\caption{Results of stellar analysis with the GSSP program and the method describe in  \citet{fridlund2010}.
For the former analysis three different normalizations of the spectrum were tested: a) without any correction, 
b) by multiplying the observed spectrum by a factor calculated from a least squares fit between
observed and synthetic spectrum, and c) with a re-normalization applied on smaller
 scales.}
\centering
\label{table::stellarmod}
\begin{tabular}{l|c|c|c|c}
\hline \hline
Parameter & \multicolumn{3}{c}{GSSP} & Method 2\\
Parameter & a) & b) & c) &\\
\hline
$T_{eff}$\,/\,K     & $ 7150  \pm 80   $  & $ 7130 \pm 80$   & $ 7350 \pm 80  $   & $ 7300 \pm 200$\\
$[Fe/H]$\,/\,dex    & $ -0.02 \pm 0.15 $  & $ -0.2 \pm 0.2$  & $ -0.15\pm 0.17$   & $ 0.0  \pm 0.2$\\
$log\,g$\,/\,cgs    & $ 3.98  \pm 0.38 $  & $ 3.96 \pm 0.34$ & $ 4.16 \pm 0.39$   & $ 4.1  \pm 0.3$\\
$vsin(i)$\,/\,km/s  & $ 127 \pm 9 $       & $ 126 \pm 10 $ & $ 130 \pm 10$ &$ 125 \pm 10 $\\
\hline
\end{tabular}
\end{table}

For the primary star we found an effective temperature $T_{eff} = (7350K \pm 250)$\,K, a
surface gravity of $log\,g = (4.16 \pm 0.39)$\,cgs, and a metalicity of $[Fe/H]
= (-0.05 \pm 0.17)$\, dex. 

The mass of the primary star $M_1$ was derived using \mbox{PARSEC1.2S}
isochrones  \citep{bressan2012, chen2014, tang2014, chen2015} in combination with
the 
stellar parameters and 2MASS color information  \citep{cutri2003}.
The radius of the primary is given by its mass and surface gravity. 
From the mass function $f(m)$ we derived the mass of the secondary object 
as $M_2 = (\Masst{}  \pm \Masste{}) M_{\sun}$. 
The radius of the secondary was calculated using the 
radius of the primary and the ratio $R_2/R_1$ that comes from the
light curve modeling $R_2 = (\Radiust{} \pm \Radiuste{}) R_{\sun}$.
The resulting system mass ($M_1+M_2$), radius of the primary ($R_1$), semi major axis $a/R_1$, and orbital period were tested for 
satisfying Kepler's third law.

We compared our results using the PARSEC1.2S model with those
using the Y2 stellar models  \citep{yi2001,demarque2004} and the Dartmouth model
 \citep{dotter2008}.
All three  models are in agreement and give us the similar results (within 1$\sigma$) for the mass and radius of the binary components. 
The Dartmouth model results in binary components that
are a bit smaller and less massive,  whereas the Y2 model 
suggests larger and more massive stars.

The atmospheric and bulk  parameters of both stars are listed in Table
\ref{table::classification}.

\begin{table}
\caption{Bulk parameters for both stars. $T_{eff}$, $Fe/H$, and $log g$ were
determined using a grid search. For the parameters derived from spectral
analysis the 1 $\sigma$ error is given. For $T_{eff}$ a 3 $\sigma$ error is
listed. Masses and radii including their errors, were determined using the
according isochrones, the results from light curve modeling, and the fitted
radial velocity measurements.}
\centering 
\label{table::classification}
\begin{tabular}{l|l}
\hline \hline
Parameter & Value \\ 
\hline
$T_{eff}$\,/\,K  & $7350 \pm 250$  \\
$[Fe/H]$\,/\,dex  & $-0.05   \pm 0.17$  \\
$log\,g$\,/\,cgs & $4.16   \pm 0.39$  \\
$v_{turb}$\,/\,km/s & $1.74^{+0.62}_{-0.41}$  \\
$vsin(i)$\,/\,km/s & $130 \pm 10 $  \\
$M_1\,/\,M_\odot $ & $\Mass{}  \pm \Masse{} $  \\
$R_1\,/\,R_\odot$ & $\Radius{}  \pm \Radiuse{} $ \\
$M_2\,/\,M_\odot$ & $\Masst{}  \pm \Masste{} $  \\
$R_2\,/\,R_\odot$ & $\Radiust{}  \pm \Radiuste{} $ \\
\hline
\end{tabular}
\end{table}

\subsection{Synchronization of the System}
\label{sec::sync}
In order to assess whether the system is synchronized we computed the
synchronization factor comparing the rotational period of the star  
with the orbital period.
If a 1:1 spin-orbit synchronization and alignment has taken place the rotation
period of the primary  star is  equal to the orbital period of the system.
We assume the orbital inclination to be nearly the same as the rotational
inclination.
The rotational velocity of the primary star derived from the spectral line
broadening is $vsini = (130 \pm 10) km/s$. We know the inclination of the orbital plane to be $i=84.1^{\circ}\pm
0.3^{\circ}$ from the light curve modeling.
With the radius of the primary star
and its real rotational velocity $V_{rot}=\frac{vsini}{sini}$ we derive the
rotational period
$P_{rot}=2*\pi*R_1/V_{rot}=(0.58 \pm 0.06)d$. This gives the synchronization
factor of
$P_{rot}/P_{orb} = 0.43 \pm 0.05 $.

The system is clearly not in a 1:1 synchronization,
but the rotational period of the primary star 
and the orbital period are close to a
 2:1 commensurability. Even if the orbital inclination would not be the same as the rotational
inclination our conclusion would still stand as the synchronization factor would only decrease for smaller inclinations.
 
Normally, we expect close binary stars to evolve into a 1:1 spin-orbit
resonance if the eccentricity is close to 0. As shown by  \citet{celletti2007, celletti2008} for examples of the solar system the 2:1 
resonances are very unlikely for objects in low eccentricity orbits.  
Our light curve and radial velocity modeling suggest an eccentricity close to 0.
This makes it unlikely for the system to be in a dynamically stable 2:1 resonance.
The observed commensurability is likely not to be a stable resonance, but a mere coincidence.
As shown in the analysis by  \citet{beky2014}, the assumption that every commensurability is due to stable dynamical resonances
is implausible.

If the binary is not yet synchronized this can only mean that it is younger than the time scale of synchronization.
This time scale for the system, $<t_{sync}>$,  was computed
according to  \citet{zahn1977} and  \citet{2001icbs.book.....H}. 
Using the stellar models grids by  \citet{claret2004}, we determined the radius
of gyrotation and the tidal torque constant of the primary star. 
For this system the time scale of synchronization lies in the range between
120\,Myrs and 250\,Myrs. If no third body is preventing the system from
synchronization, this system looks younger than 250\,Myrs.

\begin{figure}
\centering
\epsscale{1}
\plotone{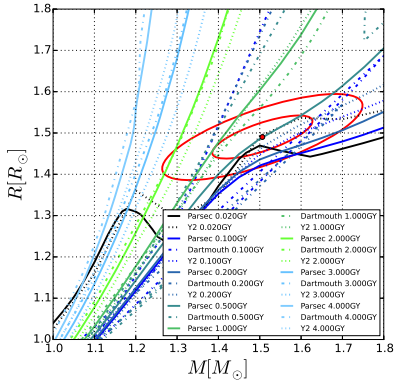}
\caption{Parsec1.2S isochrones for different ages are plotted with
continuous lines. Y2 isochrones are plotted with dotted lines. The Dartmouth model is plotted
with dashed lines. Only isochrones with solar metallicity are displayed. 
  The primary star is shown by the red marker, circles represent $1\sigma$ and $2\sigma$ error
bars.}
\label{fig::iso}
\end{figure}

For the age of the primary star we get no conclusive result, but  
Parsec1.2S isochrones suggest ages below 1.4\,Gyr.
In Figure \ref{fig::iso} the mass and radius of the primary star is plotted
along with various isochrones.

\section{Discussion}
\label{sect:summary}

The mass-radius relations given by the stellar evolution models of \citet{baraffe1998} and \citet{bressan2012}, 
indicate that the low mass companion has an   inflated radius.
The empirical mass-radius relations of   \citet[][]{mann2015} and  \citet{boyajian2012} suggests that stellar evolution models
 systematically underestimate the stellar radius of very low-mass stars by $\sim5\%$.
 For VLMSs with masses below 0.3$M_{\sun}$ the data presented in \citet[][]{mann2015} also shows discrepancy in the mass by $\sim4\%$ compared to the Dartmouth model.
However, \citet[][]{mann2015} suggest that the model inferred masses are more reliable than the empirically derived ones.
It thus is more suited to compare our results with model isochrones that are corrected for the underestimated radius.
These corrected isochrones show that the M dwarf is slightly inflated.
Such an anomalous radius could be explained by the youth of the star.

\begin{figure}
\centering
\epsscale{1}
\plotone{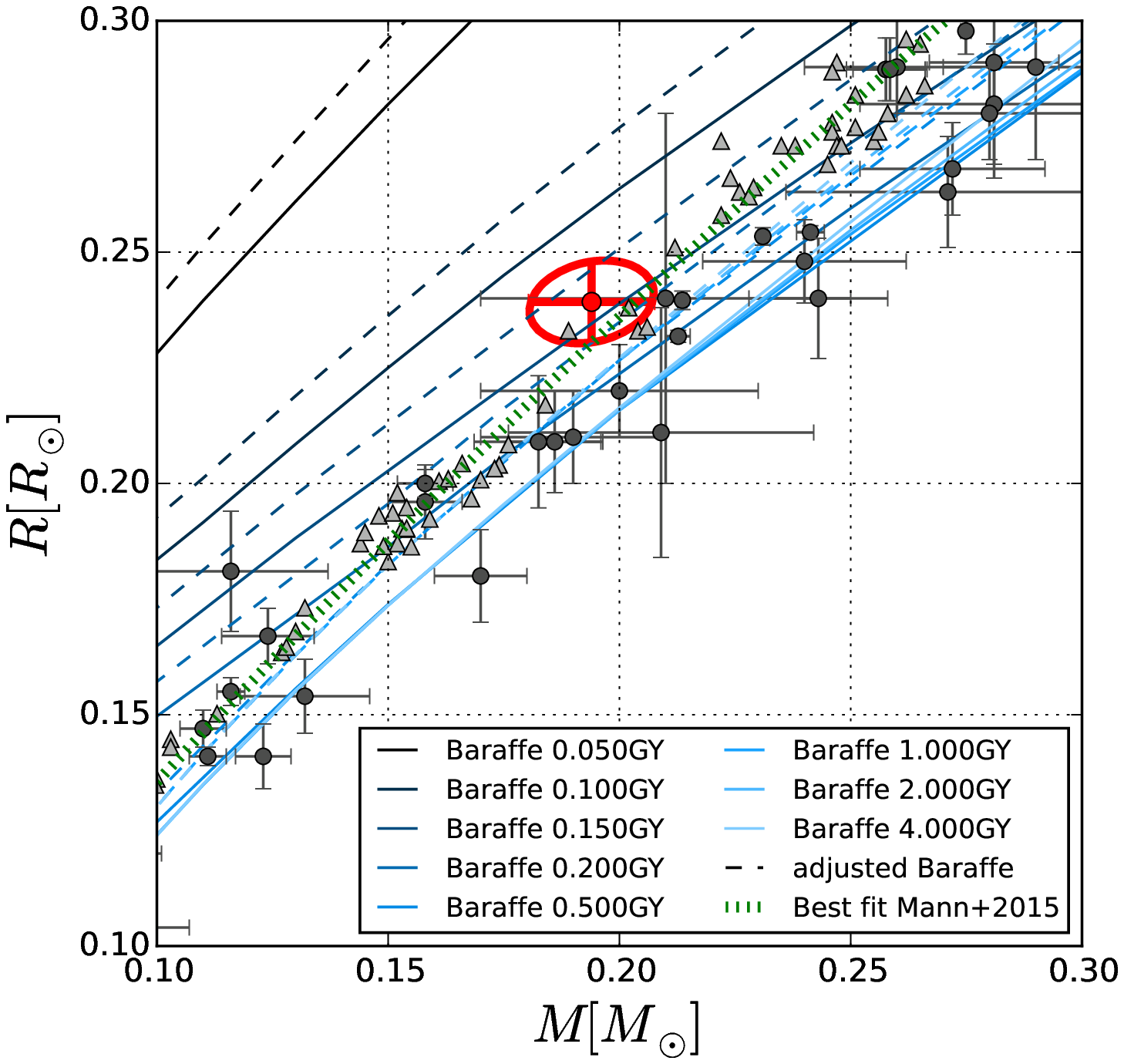}
\caption{Mass radius relation of  very low-mass stars. Plotted are stars in eclipsing binaries (gray circles) 
  \citep{segransan2003, bouchy2005, pont2005, hebb2006, pont2006, beatty2007, maxted2007, blake2008, fernandez2009, morales2009, vida2009, dimitrov2010, parsons2010, hartmann2011, irwin2011, carter2011, parsons2012, pyrzas2012,nefs2013, talor2013, gomez2014, zhou2014} and
  spectroscopic characterized single low mass stars (gray triangles) \citep{mann2015}.
  The best fit to data from \citet{mann2015} is given by the green dashed line. 
  The data is over plotted with isochrones for different ages. Continuous lines show isochrones by  \citet{baraffe1998} and the dashed lines 
  the same isochrones corrected for an radius underestimation of 5\%. 
  In red the characterized M dwarf companion with the according 1$\sigma$ error is shown.}
\label{fig::mr}
\end{figure}

Figure \ref{fig::mr} shows our M  dwarf in relation to other known
systems with masses and radii below $0.3M_{\sun}$ and $0.3R_{\sun}$, respectively.
Crosses represent eclipsing binaries and single stars studied with interferometry. Circles represent spectroscopically characterized VLMSs. 
The lines show isochrones by \citet{baraffe1998} with metallicity $[M/H]=0.0$ of
different ages. 
The dashed lines show the isochrones corrected for a radius underestimated by $5\%$.
The green dashed line shows a polynomial fit of third order to the mass-radius relation for the data presented in \citet{mann2015}.
Discrepancies between the empirical data from \citet{mann2015} and the adjusted isochrones are due to the underestimate in masses for VLMSs.
The empirical mass-radius relation for low-mass stars is based on objects typically of several Gyrs in age.
Due to the limited number of young VLMSs it is not clear whether stellar models also underestimate the radius by 5\% for young stars.
Nevertheless, taking into account the underestimate  in  the radius by the stellar models as it is known for older stars, 
the mass-radius relation of the M dwarf agrees best with the isochrones for ages between 100\,Myrs\,-\,200\,Myrs.

Comparison of the stellar parameters for the primary star with
isochrones do not allow us to constrain further the age of the system,
but our results hint towards a young system.
Isochrones from different stellar models all suggest an age 
below 1\,Gyr. Furthermore, the system is not in a 1:1
spin orbit resonance, which we would expect for such binary system with an eccentricity close to 0.

The stellar rotation of the primary star is close to  a 2:1 commensurability
with the orbital period. 
Similar commensurabilities were found in some exoplanetary systems
 \citep[see][]{beky2014} and in the brown dwarf system CoRoT-33
 \citep{csizmadia2015}, but have not yet been reported for binary systems.\\
It is unlikely that these 2:1 resonant  systems of low eccentricities are dynamically stable \citep{celletti2008}. 
As pointed out in the study by  \citet{beky2014} there are good reasons to believe that such commensurabilities are 
a statistical phenomena and not a stable resonance.

We see two possibilities why this system is not tidally locked. 
Either the system is younger than the time scale of synchronization, which is
below 250\,Myrs,  or
a third body is present that perturbs the system. 
However, we find no evidence for this third body
in the photometric or RV data. Long-term high precision RV monitoring, or AO imaging of this
star may reveal a third body.
At the present time, all the available evidence from
the dynamical analysis of of the system combined with the
mass-radius relationship of both components point to a system
that is younger than 250\,Myrs.

In contrast to M dwarfs older than 500\,Myrs, where the differences between
stellar evolution models are small compared to observational errors,
isochrones of ages below 250\,Myrs differ significantly between
models.
Given the uncertainties in the stellar parameters it is not yet possible to
distinguish between different stellar models for this M dwarf.
But with the expected age of the system below the time scale of synchronization,
which is in agreement with the mass-radius relation of the low mass companion,
this system is a unique test object for stellar evolution models.
It is one of the youngest studied M dwarfs in an eclipsing binary.
Better values of the stellar parameters, particularly the stellar age of the
primary star, will allow to test different stellar evolution models.
Additionally this system can serve as an interesting test object for rotational
evolution of low-mass stars in presence of a close companion and possibly strong
stellar wind  \citep[c.f.][]{ferraz2015}.\\

\section{Conclusion}

We characterized a detached eclipsing binary system with un-equal mass components
comprised of a very low-mass M dwarf orbiting an early
 F-type main-sequence star.
The system was investigated combining photometric data and radial velocity measurements. Using stellar evolution models we determined the bulk properties
 of the primary star.
Using different stellar models for the characterization of the primary star did not lead to significant changes in the mass-radius relation of either of the stars.

The orbital period is $1.35121 \pm 0.00001$ days. The mass of the M dwarf is $M_2= \Masst{} \pm \Masste{} M_\sun$. 
With a   radius of $R_2 = \Radiust{} \pm \Radiuste{} R_\sun$ the M dwarf is slightly inflated even when
taking into account that current stellar models underestimate the radii of low-mass stars by $\sim 5\%$.

The low density of the M dwarf star could be explained by an age of the system between 100\,Myrs and 250\,Myrs. 
The spectral characterization of the primary star does not allow us to further constrain the age of the system.
However, the system has not yet reached the 1:1 spin-orbit resonance, which we would expect for such a close binary 
 with a nearly circular orbit. This supports the conclusion that the age of the system is below 250\,Myrs.

The M dwarf thus is one of the youngest characterized main-sequence M dwarfs in an eclipsing binary system.
Additionally, it is one of the very few VLMSs which allows us to estimate the age estimate without isochrone fitting.
It might play a crucial role in further understanding of the mass-radius relation for young very low mass objects.     
The system is also of high interest with regard to the dynamical interactions in such close binaries.

\acknowledgments
Part of this work was supported by the
Deut\-sche For\-schungs\-ge\-mein\-schaft DFG\/ under projects
Ei409/14-1,-2. Sz.Cs. acknowledges the support under the Hungarian OTKA Grant
K113117. IRAF is distributed by the National Optical Astronomy Observatory,
which is operated by the Association of Universities for Research in Astronomy
(AURA) under cooperative agreement with the National Science Foundation. PyRAF
and PyFITS were used for this work and are products of the Space Telescope
Science Institute, which is operated by AURA for NASA. 
This work is based in part on observations obtained with the 2-m Alfred Jensch
Telescope of the Th\"uringer Landessternwarte Tautenburg. 
We thank the referee for the detailed comments which have improved this paper.

{\it Facility:} \facility{TLS (2m Coude, TEST)}

\label{sec:bib}

\end{document}